\documentclass[letterpaper, 10 pt, conference]{ieeeconf}  % Comment this line out if you need a4paper

\IEEEoverridecommandlockouts                              % This command is only needed if 
\usepackage{graphicx} % for pdf, bitmapped graphics files
\usepackage{algorithmic}
\usepackage{booktabs}
\usepackage{wrapfig}
\usepackage{times} % assumes new font selection scheme installed
\usepackage{amsmath} % assumes amsmath package installed
\DeclareMathOperator{\Tr}{Tr}
\usepackage{amssymb}  % assumes amsmath package installed

\title{\LARGE \bf
Online Policies for Real-Time Control Using MRAC-RL}

\author{Anubhav Guha and Anuradha M. Annaswamy% <-this % stops a space
\thanks{This work is supported by the Boeing Strategic University Initiative.}% <-this % stops a space
\thanks{A. Guha (anguha@mit.edu) and A.M. Annaswamy (aanna@mit.edu) are with the Department of Mechanical Engineering, Massachusetts Institute of Technology, Cambridge, MA, 02139 USA. }%
}
\begin{document}

\maketitle
\thispagestyle{empty}
\pagestyle{empty}

%%%%%%%%%%%%%%%%%%%%%%%%%%%%%%%%%%%%%%%%%%%%%%%%%%%%%%%%%%%%%%%%%%%%%%%%%%%%%%%%
\begin{abstract}
In this paper, we propose the Model Reference Adaptive Control \& Reinforcement Learning (MRAC-RL) approach to developing online policies for systems in which modeling errors occur in real-time. Although reinforcement learning (RL) algorithms have been successfully used to develop control policies for dynamical systems, discrepancies between simulated dynamics and the true target dynamics can cause trained policies to fail to generalize and adapt appropriately when deployed in the real-world. The MRAC-RL framework generates online policies by utilizing an inner-loop adaptive controller together with a simulation-trained outer-loop RL policy. This structure allows MRAC-RL to adapt and operate effectively in a target environment, even when parametric uncertainties exists. We propose a set of novel MRAC algorithms, apply them to a class of nonlinear systems, derive the associated control laws, provide stability guarantees for the resulting closed-loop system, and show that the adaptive tracking objective is achieved. Using a simulation study of an automated quadrotor landing task, we demonstrate that the MRAC-RL approach improves upon state-of-the-art RL algorithms and techniques through the generation of online policies.
\end{abstract}

%%%%%%%%%%%%%%%%%%%%%%%%%%%%%%%%%%%%%%%%%%%%%%%%%%%%%%%%%%%%%%%%%%%%%%%%%%%%%%%%
\section{INTRODUCTION}

\label{sec1}
Recent years have witnessed an explosive growth in the field of reinforcement learning (RL) and its use for the development of control policies for complex systems and environments. Successful applications have been broad and varied - ranging from direct actuator-level control and state regulation to high-level planning and decision making \cite{mnih2015}\cite{Ng2006}\cite{lillicrap2019continuous}\cite{kober2013}. The effectiveness of reinforcement learning algorithms in overcoming constraints that typically limit classical control techniques has enabled RL's application to decision making and continuous control tasks \cite{recht2018tour}\cite{schulman2018highdimensional}.

Many RL algorithms are fundamentally data-driven methods. As a result, control polices are often learned largely in simulation. Training in simulation is a powerful technique, allowing for a near infinite number of agent-environment interactions - in comparison, training a policy on an actual plant could be expensive, time-consuming or dangerous. In practice, however, offline policies trained in simulation often exhibit degenerate performance when used for real-time control due to modeling errors that can occur online \cite{Koos2010}\cite{tan2018simtoreal}. It may be difficult to reliably predict the behavior of a learned policy when it is applied to an environment different from the one seen during training \cite{fulton2018safe}\cite{AurkoMismatch}\cite{rajeswaran2018generalization}\cite{packer2019assessing}. As a result, many researchers have focused on methods to bridge this so-called "sim-to-real" gap.

In this paper we introduce a framework that leads to an online policy that can be applied to the control of systems when modeling errors occur online. This policy combines Model Reference Adaptive Control (MRAC) and RL, which we denote as MRAC-RL, and forms the main contribution of the paper. The MRAC-RL control architecture includes an inner-loop and an outer-loop, with adaptive control elements in the inner-loop and RL elements in the outer-loop. This architecture allows adaptive algorithms to adjust outer-loop commands online, so that modeling errors due to parametric uncertainties can be accounted for. The central merit of this MRAC-RL framework is that it enables the true system to react to the learned control policy in the same way that the simulated system would have responded during training when no modeling errors were present.
\subsection{Related Work}\label{sec:1.1}
\subsubsection{Reinforcement Learning}\label{sec:1.1.1}
A number of reinforcement learning algorithms have been successfully used to solve continuous control tasks. Of these, the Proximal Policy Optimization (PPO) algorithm \cite{schulman2017proximal} has been shown to be successful in a variety of tasks \cite{wang2019benchmarking}\cite{duan2016benchmarking}\cite{henderson2019deep}. Even the most powerful RL algorithms, however, may fail to generalize in the presence of modeling errors \cite{higgins2018darla}\cite{nagabandi2019learning}\cite{lake2016building}. The research community has largely tackled these challenges by developing specialized RL algorithms. For example, the model-based PILCO \cite{Deisenroth2011} uses a learned probabilistic dynamics model to account for dynamic uncertainty, while DARLA \cite{higgins2018darla} improves sim-to-real transfer by learning robust features. Another popular approach is to directly modify the simulation \& training protocols. In \cite{rajeswaran2017epopt} an ensemble of environments with varying dynamics were used to improve the robustness of learned policies, while \cite{loquercio2020drone} used simulated domain randomization to bridge the sim-to-real gap on a drone racing task.

As alluded to above, much of the research in bridging the sim-to-real gap has focused on improved simulation techniques and improved RL algorithms \cite{pinto2017robust}\cite{packer2019assessing}\cite{higgins2018darla}\cite{Deisenroth2011}\cite{nagabandi2019learning}\cite{rajeswaran2018generalization}\cite{berkenkamp2017safe}. There has, however, been little attention paid to methods that may be used to inject additional robustness and adaptability into an already-trained policy.
\subsubsection{Adaptive Control}
Adaptive control and system identification methods have long been used in the control of safety-critical systems \cite{L1AC}\cite{dydek2013}\cite{WISE2011}\cite{michini2009}\cite{Wiese2013}. Unlike many RL algorithms, adaptive control techniques excel in the "zero-shot" enforcement of control objectives - that is, in learning to accomplish a task online \cite{recht2018tour}\cite{narendra1989}\cite{aastrom2013adaptive}\cite{ioannou2012robust}. These adaptive techniques are able to accommodate, in real-time, parametric uncertainties and constraints on the control input magnitude \cite{Karason1994}\cite{Lavretsky2004} and rate \cite{gaudio2019adaptive}. This ability to achieve control goals while accounting for parametric uncertainties in real-time is the strength of adaptive control.

The approach that we propose in this paper is a combination of the MRAC and RL methods so as to realize their individual advantages and minimize their weaknesses. While MRAC approaches are able to accommodate the presence of modeling errors over short time-scales and meet tracking objectives, they are unable to guarantee the realization of long-term optimality-based objectives. In contrast, RL-trained policies can handle a broad range of tasks \& objectives \cite{sutton}, but may fail to generalize appropriately in the presence of modeling errors (as discussed in Section \ref{sec:1.1.1}). The MRAC-RL controller that we propose in this paper combines these two methods, with MRAC in the inner-loop and RL in the outer-loop. The resulting controller is guaranteed to be stable under the assumption that the outer-loop has been designed, offline, to lead to optimal behavior when no modeling errors exist.

The general problem of interest is posed in Section \ref{sec:2}. A brief description of the RL approach is also included in this section. The MRAC-RL architecture and stability results are presented in Section \ref{sec:3}. An extensive numerical study is reported in  Section \ref{sec:4}, in which a quadrotor subjected to modeling errors and loss of effectiveness is tasked with landing on a moving platform. In these experiments, the performance of MRAC-RL is validated and compared with standard approaches. Summary and conclusions are presented in Section \ref{sec: 5}.

\section{PROBLEM STATEMENT}\label{sec:2}
Consider a continuous-time, deterministic nonlinear system described by the following dynamics:
\begin{equation}\label{eq:dynamics}
\begin{split}
    &\dot{x}_1 = \delta_1x_2\\
    &\dot{x}_2 = \delta_2x_3\\
    &\quad \dots\\
    &\dot{x}_{n-1} = \delta_{n-1}x_n\\
    &\dot{x}_n = \alpha^T\zeta(x) + bu 
\end{split}
\end{equation}
where $x$ is the $n-$dimensional state vector, and $\zeta(x) = [\phi_1(x), \phi_2(x), \dots, \phi_n(x)]^T$ denotes system nonlinearities. $\nu_i$ denotes the $i$th component of a vector $\nu$. We compactly represent (\ref{eq:dynamics}) as:
\begin{equation}\label{eq:PCC}
\begin{aligned}
\dot{x}_i(t) = \begin{cases} 
      \delta_i x_{i+1}(t) & i \neq n \\
      \alpha^T \zeta(x(t)) + bu(t) & i = n
   \end{cases}
\end{aligned}
\end{equation}
The problem that we address is the determination of the control input $u(t)$ in (\ref{eq:PCC}) in real-time, when $\delta$ and the nonlinearity $\zeta(\cdot)$ are known, but the parameters $\alpha$ and $b$ are unknown. The goal is to choose $u$ so as to minimize a cost function of $x$ and $u$. Compactly, this is formulated as the following: 

\begin{equation}\label{eq:optimal_control}
\begin{aligned}
\min_{{\scriptscriptstyle u(t) \in U \; \forall t \in [0, T]}} \quad & \int_{0}^{T}c(x(t), u(t))dt\\
\textrm{subject to} \quad & \textrm{Dynamics in (\ref{eq:PCC}) } \forall t \in [0, T]\\
  &x(0) = x_0   \\
\end{aligned}
\end{equation}
for a given initial condition $x_0$. As a matter of notational convenience, we often suppress signal dependencies on the time variable $t$ (for example, $u(t)$ may be referred to by $u$). Furthermore, a shorthand is adopted in which we use $\zeta$ to mean $\zeta(x(t))$.
\subsection{An Offline Approach Based on Reinforcement Learning}
The presence of unknown parameters $\alpha$ and $b$ introduces a direct challenge in determining an RL-based solution, as a simulation cannot be constructed for training purposes. To overcome this, we start with nominal, known parameter values $\alpha_r$ and $b_r$ for $\alpha$ and $b$, respectively. Using these values we formulate the following reference model:
\begin{equation}\label{eq:PCC_lambda_r}
\begin{aligned}
\dot{x}_{r, i}(t) = \begin{cases} 
      \delta_i x_{r, i+1}(t) & i \neq n \\
      \alpha_r^T \zeta(x_r(t)) + b_ru_r(t) & i = n
   \end{cases}
\end{aligned}
\end{equation}
As (4) is fully known, one can employ a number of approaches in order to choose the reference-control input $u_r(t)$ so that the accumulated cost in (\ref{eq:optimal_control}) can be minimized. The argument therefore is that if the true parameters in (\ref{eq:PCC}) are equal to their nominal values, then a choice of $u(t)=u_r(t)$ will solve the problem in (\ref{eq:optimal_control}).
The determination of $u_r(t)$ can be carried out using a number of methods, including LQR and MPC, with the former providing optimal solutions for linear dynamic systems with quadratic costs, and the latter when the control input is subjected to various constraints.

An alternate offline approach that can be used to solve this problem is reinforcement learning. RL generates a feedback policy $\pi$. If the RL training has been successful, this choice of control will drive the system in (\ref{eq:PCC_lambda_r}) so that the accumulated cost in (\ref{eq:optimal_control}) is minimized.

In order to lead to a self-contained exposition, we briefly describe the RL training procedure. First, it is assumed that the continuous time dynamics in (\ref{eq:PCC_lambda_r}) are sampled with sufficient accuracy, resulting in the discrete time dynamics:

\begin{equation}\label{eq:tran}
\begin{aligned}
x_{r, t+1} \sim p(x_{r, t+1} | x_{r, t}, u_{r, t})
\end{aligned}
\end{equation}

\noindent An appropriate numerical integration scheme ensures that this discrete-time formulation closely approximates the dynamics in (\ref{eq:PCC_lambda_r}). The resulting Markov Decision Process (MDP) is then utilized to lead to a policy $\pi$ such that $u_r=\pi(x_r)$, where $\pi(\cdot)$ is stochastic. RL begins with the construction of a simulation environment (using the reference model) that is used to collect the system data for an initial policy. Repeated training of the policy $\pi$ is then carried out, so as to optimize the cost function \cite{kaelbling1996reinforcement}. 

The RL training procedure consists of repeated interactions of the policy with the environment as follows. At each timestep, an observation $x_{r, t}$ is received, a control $u_{r,t}$ is chosen, and the resulting cost $c$ is received. Repeating this process, a set of input-state-cost tuples $\mathcal{D} = [(x_{r, 1},u_{r, 1}, c_1),\ldots (x_{r, N},u_{r, N}, c_N)]$ are formed. These tuples form the data used to train the reinforcement learning agent. This data is then used to update the policy using an RL algorithm. In typical RL algorithms the policy is parametrized by a set of parameters, so that $u_t \sim \pi_\theta (x_t)$. For example, in deep RL $\theta$ represents the weights/biases of a neural network. The learning algorithm then seeks to adjust $\theta$ so that the expected accumulated cost is minimized, i.e
\begin{equation}\label{eq:reward}
\begin{aligned}
\min_{\theta} \quad & J(\theta)\\
\textrm{where} \quad & J(\theta) = \mathbb{E}_{\pi_\theta}\left[\sum_{t = 0}^T c_t \right ]
\end{aligned}
\end{equation}
The expectation arises from the potential stochasticity of the policy $\pi$ and the transition dynamics \cite{schulman2015high}\cite{vidyasagar2020recent}. A large number of approaches to solving (\ref{eq:reward}) exist, including policy gradient methods in which we attempt to directly estimate $\nabla_\theta J(\theta)$. The policy gradient can be calculated as:
\begin{equation}\label{eq:policy_gradient}
    \nabla_\theta J(\theta)  = \mathbb{E} \left [-\sum_{t = 0}^T A^{\pi_\theta}(x_t, u_t) \nabla_\theta \log \pi_\theta (u_t | x_t)\right ]
\end{equation} where $A^{\pi_\theta}$ is the advantage function \cite{schulman2015high}, which represents the relative utility of taking an action $u_t$ in state $x_t$, with respect to the other potential viable actions. The gradient in (\ref{eq:policy_gradient}) is then utilized to update the weights as $\theta_{t+1} = \theta_{t} - \eta \nabla_{\theta_{t}} J(\theta_{t})$ with learning rate $\eta$. With this update, the policy $\pi$ is updated, new data $\mathcal{D}$ is collected and the updated policy gradient (\ref{eq:policy_gradient}) is calculated. The process repeats until training is complete. For a more detailed overview on policy gradient algorithms, refer to \cite{schulman2015high}.

In some simple settings, certain reinforcement learning algorithms can be shown to converge to optimal policies. For example, if the state and action spaces are discrete (that is, $x_t \in \mathbb{X}$, $u_t \in \mathbb{U}$ $\forall t$ for finite sets $\mathbb{X}, \; \mathbb{U}$) Watkin's $Q$-learning algorithm can be used to achieve the globally optimal policy. The optimal $Q$ function, or state-action value function, is given by the fixed point solution to the Bellman equation:
\begin{equation}\label{eq:bellman}
    Q^*(x, u) = -c(x, u) + \gamma \sum_{x' \in \mathbb {X}} p(x' | x, u) \max_{u' \in \mathbb{U}} Q^*(x', a')
\end{equation}
Here, $0 < \gamma < 1$ is a discount factor. If the optimal $Q$ function is found, the globally optimal policy can easily be determined: $\pi^*(x) =$ arg$\max_{u \in \mathbb{U}} Q^*(x, u)$. An iterative rule for determining the optimal $Q$-function, denoted as $Q$-learning, is given by:
\begin{equation}\label{eq:q_update}
\hat{Q}(x_t, u_t) = \hat{Q}(x_t, u_t) + \eta_t \delta_t\\
\end{equation}

\noindent where $\eta_t$ is a sequence of learning rates, and $\delta_t = -c_t + \gamma \max_{u' \in \mathbb{U}} \hat{Q}(x_{t+1}, u') - \hat{Q}(x_t, u_t)$ is the temporal difference error \cite{watkins1992q}\cite{vidyasagar2020recent}. If $\eta_t$ satisfy the Robbins-Monro criterion and every state-action pair (as represented in the trajectory) is visited infinitely often, then the $Q$ function estimate can be shown to converge to the fixed-point solution: $\hat{Q}(x, u) \rightarrow Q^*(x, u) \; \forall x \in \mathbb{X}, u \in \mathbb{U}$ \cite{watkins1992q}.

Two important points should be made about these guarantees of convergence. First, it is assumed that the transition probabilities in (\ref{eq:tran}) do not change during the $Q$-learning process. Therefore, that (\ref{eq:tran}) is fixed is a requirement for the convergence result. The second point is the requirement that $\mathbb{X}$ and $\mathbb{U}$ are finite sets. In the context of controlling a physical system, representing the state and action spaces as continuous is often a more appropriate modeling choice. The MRAC-RL approach proposed in this paper attempts to relax both of these requirements. Adaptive control is used to accommodate changing transition/dynamic models, while policy gradient algorithms such as PPO are used at an outer-loop to accommodate continuous state and action spaces. It should be noted that a significant and growing body of literature exists in RL, including on methods that attempt to relax the aforementioned requirements \cite{devraj2017zap}\cite{berkenkamp2017safe}\cite{loquercio2020drone}. The approach that we propose, MRAC-RL, is distinct from these and it's details and advantages will be presented in the sections that follow.

As mentioned earlier, the RL literature is large and growing. In this paper we restrict our attention to the Proximal Policy Optimization algorithm in \cite{schulman2017proximal}. PPO is a policy gradient actor-critic algorithm which attempts to solve (\ref{eq:reward}). Deep actor-critic RL methods employ multiple neural networks, so that the policy and value functions are learned simultaneously.

A specific point to note about the proposed MRAC-RL is its deterministic nature. We assume that the underlying model is deterministic, and propose a corresponding deterministic control solution. A stochastic policy trained via RL can also be made deterministic by the choice of a random seed. With this solution as the first-step, subsequent extensions to its stochastic counterpart need to be carried out, which are beyond the scope of this paper.
\subsection{Modeling Uncertainty}
Following the approach above, RL is used to train a feedback policy $\pi$ for the reference system in (\ref{eq:PCC_lambda_r}), so that $u_{r, t} = \pi(x_{r, t})$. If the true parameters $\alpha$ and $b$ coincide with their nominal values $\alpha_r$ and $b_r$, respectively, then the reference model in (\ref{eq:PCC_lambda_r}) is identical to the true system in (\ref{eq:PCC}). Since, under this assumption, the target system is identical to the reference system, a choice of $u_t=u_{r, t} = \pi (x_{r, t})$ guarantees that the same pseudo-optimal performance exhibited in the reference system can be assured in the target system.

The problem that we consider in this paper is the case when the true parameters depart from their nominal values. As this departure is assumed to occur in real-time, it cannot be accommodated for in the training procedure. In such a case, the target transition function differs from the reference system transition function, and therefore the policy $\pi$ cannot be guaranteed to converge, i.e., the training is incomplete. To accommodate this real-time change, a faster feedback loop based on adaptive control is proposed. The details of the resulting MRAC-RL architecture are described in Section III.

\subsection{Example}\label{sec:example}
We illustrate the problem statement using a simple example: the control of an inverted pendulum, whose model is given by (\ref{eq:pendulum}):

\begin{equation}\label{eq:pendulum}
    \begin{aligned}
        ml^2\ddot{\theta} = mgl\sin\theta -\mu\dot{\theta} + u
    \end{aligned}
\end{equation}

\noindent It is easy to see that (\ref{eq:pendulum}) corresponds to a special case of (\ref{eq:PCC}). The problem that we address is the control of (\ref{eq:pendulum}) in the presence of uncertainties in $m,l,\mu$, with only their nominal values $m_r$, $l_r$, and $\mu_r$ known. That is, the parameters in (\ref{eq:PCC}) correspond to $\delta_1=1$, $\alpha=[g/l, -\mu/ml^2]^T$, and $b=1/(ml^2)$, with nominal values $\alpha_r$ and $b_r$ defining the reference system in (\ref{eq:PCC_lambda_r}). With no parametric uncertainty, it is clear that RL approaches can be used in order to determine $u$ so that a requisite objective in (\ref{eq:optimal_control}) can be minimized \cite{duan2016benchmarking}. The goal is to determine the policy for $u$ in (10) when $\alpha$ and $b$ depart from their nominal values.

\section{An ONLINE POLICY BASED ON MODEL REFERENCE ADAPTIVE CONTROL}\label{sec:3}
\begin{wrapfigure}{L}{0.3\textwidth}
\centering
\includegraphics[width=0.25\textwidth]{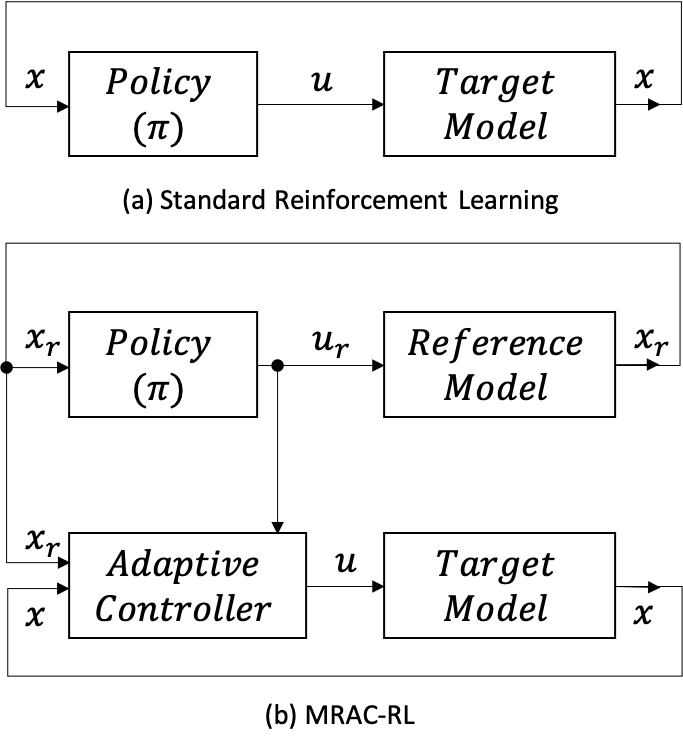}
\caption{\label{fig:mracrl}RL vs. MRAC-RL. (a) represents a standard application of a trained policy, in which the trained policy is inserted directly into the target system: $u = \pi(x)$. (b) shows how MRAC-RL is used. The policy is inserted into the reference system, producing $u_r = \pi(x_r)$. (\ref{eq:xi}) - (\ref{eq:adaptive2}) are then used to calculate $u$.}
\end{wrapfigure}

The problem that we address is the control of (\ref{eq:PCC}) when $\alpha$ and $b$ are unknown. For ease of exposition, we express $b=\lambda b_r$ and assume that $\lambda >0$. We make the following assumption about the reference system (\ref{eq:PCC_lambda_r}), which is fully known:

\noindent \textit{Assumption A1:}  An RL-based solution $u_r(t)$ can be determined for $t\geq 0$ as the approximate solution of the problem:

\begin{equation}\label{eq:optimal_control_r}
\begin{aligned}
\min_{{\scriptscriptstyle u_r(t) \in U \; \forall t \in [0, T]}} \quad & \int_{0}^{T}c(x_r(t), u_r(t))dt\\
\textrm{subject to} \quad & \textrm{Dynamics in (\ref{eq:PCC_lambda_r}) } \forall t \in [0, T]\\
  &x_r(0) = x_0   \\
\end{aligned}
\end{equation}

\noindent for all initial conditions $x_0$. 

In what follows, we denote $e_x(t)=x(t)-x_r(t)$ and $e_\zeta(t) = \zeta (x(t)) - \zeta(x_r(t))$. Before presenting the online policy, we introduce a few notations and definitions. We consider the matrix $A_H$ in the following form:

\begin{equation}\label{eq:AH}
\begin{aligned}
A_H = \begin{bmatrix}
        0 & \delta_{1} & 0 & \dots & 0\\
        0 & 0 & \delta_{2} & \dots & 0\\
        \vdots & \vdots & \vdots & \ddots & \vdots\\
        0 & 0 & 0 & \dots & \delta_{n-1}\\ 
        \alpha_{H, 1} & \alpha_{H, 2} & \alpha_{H, 3} & \dots & \alpha_{H, n}
        \end{bmatrix}
\end{aligned}
\end{equation}
where $\delta$ is the fixed vector corresponding to $\delta$ in (\ref{eq:PCC}) and (\ref{eq:PCC_lambda_r}). $\alpha_H$ is chosen so that the characteristic polynomial of $A_H$ has eigenvalues $\sigma_i$ such that $Re[\sigma_i] < 0$ $\forall$ $i=1, \dots, n$. The choice of these $\sigma_i$ represent hyperparameters, and render $A_H$ Hurwitz. It is easy to see that there always exists $\alpha_H$ that will ensure such a spectrum for $A_H$. We now choose the MRAC-RL input signal $u(t)$ in (\ref{eq:PCC}) as follows (see Figure \ref{fig:mracrl}):
\begin{equation}\label{eq:xi}
\begin{aligned}
\xi = u_r - \frac{1}{b_r}\alpha_r^Te_\zeta + \frac{1}{b_r}\alpha_H^Te_x
\end{aligned}
\end{equation}

\begin{equation}\label{eq:adaptive1}
    \begin{aligned}
        u = \hat{K}_\zeta^T\zeta + \hat{k}_\xi\xi
\end{aligned}
\end{equation}
\begin{equation}\label{eq:adaptive2}
    \begin{aligned}
    \dot{\hat{K}}_\zeta = -\Gamma_\zeta \zeta e_x^TPB_r, \quad \dot{\hat{k}}_\xi = -\gamma_\xi \xi e_x^TPB_r
\end{aligned}
\end{equation}

\noindent where $\Gamma_\zeta = \Gamma_\zeta^T \succ 0$, $\gamma_\xi > 0$ and $P = P^T \succ 0$ solves the Lyapunov equation: $PA_H + A_H^TP = -Q$ for a given symmetric positive definite matrix $Q$. $\hat{K}_\zeta$ and $\hat{k}_\xi$ represent adaptive parameter estimates that are updated at every timestep. It is clear from (\ref{eq:xi}) - (\ref{eq:adaptive2}) that if there are no parametric uncertainties, if the initial conditions of (\ref{eq:PCC}) are identical to those of (\ref{eq:PCC_lambda_r}), if $\hat K_\zeta(0)=0_{n \times 1}$ and if $\hat{k}_\xi(0)=1$, then the MRAC-RL policy coincides with $u_r(t)$. Assumption A1 then implies that the optimization problem in (\ref{eq:optimal_control}) is (approximately) solved. The following theorem articulates the stability property of the MRAC-RL architecture when there is parametric uncertainty in (2):\\\\
\textit{Theorem 1:}
    Under Assumption A1, the closed-looped systems specified by the target system (\ref{eq:PCC}), the reference system (\ref{eq:PCC_lambda_r}), and the MRAC-RL controller given by (\ref{eq:adaptive1}) - (\ref{eq:adaptive2}) will have globally bounded solutions, with $\lim_{t \to \infty} ||e_x(t)|| = 0$.
\\\\
\textit{Proof:}
In order to determine the underlying Lyapunov function for the closed loop system, we first define the ideal control parameters for the MRAC-RL architecture. In particular, define $K_\zeta^*$ and $k_\xi^*$ as the solutions of the so-called matching conditions:

\begin{equation}\label{eq:match}
\begin{aligned}
\alpha+\lambda b_r K_\zeta^* = \alpha_r, \quad \lambda k_\xi^* b_r = b_r
\end{aligned}
\end{equation}
Inserting the adaptive control law (\ref{eq:adaptive1}) into the closed-loop target system, we next determine the dynamics of the state tracking error, $e_x$. For $i = 1, \dots, n-1$ it can be seen that $\dot{e}_{x_i} = \delta_i e_{x_{i+1}}$. The derivative of $e_{x_n}$ can then be determined as: 
\begin{flalign*}
\dot{e}_{x_n} &= \alpha^T\zeta - \alpha^T_r \zeta_r - b_r u_r&&\\
&+ \lambda b_r \hat{K}_\zeta^T\zeta + \lambda b_r \hat{k}_\xi u_r - \lambda \hat{k}_\xi \alpha^T_r e_\zeta + \lambda \hat{k}_\xi \alpha^T_H e_x
\end{flalign*}
Utilizing the matching conditions (\ref{eq:match}) and defining the parameter estimation errors $\tilde{K}_\zeta = \hat{K}_\zeta - K^*_\zeta$, $\tilde{k}_\xi = \hat{k}_\xi - k^*_\xi$, we can rewrite this as: $\dot{e}_{x_n} = \alpha_H^Te_x + \lambda b_r [ \tilde{K}_\zeta^T \zeta + \tilde{k}_\xi\xi]$. Defining $B_r = [0, 0, \dots, b_r]^T$, the complete error equation can be written as:
\begin{equation}
    \begin{aligned}
    \dot{e}_x = A_H e_x + \lambda B_r [ \tilde{K}_\zeta^T \zeta + \tilde{k}_\xi\xi]
    \end{aligned}
\end{equation}
This leads to a Lyapunov function candidate:
\begin{equation}\label{eq:lyapunov}
    V(e_x, \tilde{K}_\zeta, \tilde{k}_\xi) = e_x^TPe_x + \lambda \Tr(\tilde{K}^T_\zeta\Gamma_\zeta^{-1}\tilde{K}_\zeta) + \lambda\frac{\Tilde{k}_\xi^2}{\gamma_\xi}
\end{equation}
The time derivative of $V$ can be calculated as:
\begin{flalign*}
\dot{V} &= -e_x^TQe_x + 2\lambda\Tr(\Tilde{K}_\zeta^T[\zeta e_x^TPB_r + \Gamma_\zeta^{-1}\dot{\hat{{K}}}_\zeta])&&\\
&+ 2\lambda \Tilde{k}_\xi[e_x^TPB_r\xi  + \frac{\dot{\hat{k}}_\xi}{\gamma_\xi}]
\end{flalign*}
Choosing the adaptive parameter update laws as (\ref{eq:adaptive1}) and (\ref{eq:adaptive2}), we therefore obtain that:
\begin{equation*}
    \dot{V} = -e_x^TQe_x \leq 0
\end{equation*}
This leads to the conclusion that $e_x(t)$, $\hat{K}_\zeta(t)$, and $\hat{k}_\xi(t)$ are bounded for all $t\geq 0$. From Assumption A1, it follows that $x_r(t)$ is bounded, and therefore it follows that the state $x(t)$, $\zeta(x(t))$, and $u(t)$ are all bounded for any initial conditions. Barbalat's lemma can be utilized to conclude that $\lim_{t \to \infty} ||e_x(t)||=0$, which concludes the proof \cite{narendra1989}. $\blacksquare$

The MRAC-RL solution that we propose in this paper is given by (\ref{eq:xi})-(\ref{eq:adaptive2}), where the MRAC component occurs in (\ref{eq:adaptive1}) and (\ref{eq:adaptive2}), and the RL component occurs in (\ref{eq:xi}). The last two terms in (\ref{eq:xi}) are essential coupling term that guarantee that the combination of MRAC and RL results in a control architecture that can be globally bounded.

The structure of the MRAC-RL controller differs from the standard MRAC solution in the following manner. MRAC is typically used to track systems in which the reference input $u_r$ has been chosen so that the reference system, which is selected to be stable, produces a desired and bounded reference signal $x_r$. In the scenario given in this paper, there is no such requirement on the reference system (\ref{eq:PCC_lambda_r}). The only requirement is that $u_r$ has been trained by the RL policy so that Assumption A1 is met. This flexibility in choosing $u_r$ allows any learning that stems from RL to be explicitly incorporated into the controller, and hence is advantageous compared to a pure MRAC solution.

The benefit of introducing MRAC into the architecture is clear  - the controller in (\ref{eq:adaptive1})-(\ref{eq:adaptive2}) guarantees in real-time that the state tracks that of the reference system, with a clear analytical guarantee of global boundedness, despite the departure of the true target system model (\ref{eq:PCC}) from the reference system (\ref{eq:PCC_lambda_r}). We have, however, not demonstrated the optimality of the cost function in (\ref{eq:optimal_control}) for the proposed control input $u(t)$. In a specific case where the cost function does not depend on $u(t)$ then MRAC-RL can be shown to approximately optimize (\ref{eq:optimal_control}) as $T \rightarrow \infty$. While analytic evaluation of MRAC-RL control optimality is a direction for future work, the simulation study in the following section demonstrates that MRAC-RL can drastically improve the control performance for a general cost function. 

\section{SIMULATION EXPERIMENTS}\label{sec:4}

We investigate a quadrotor task, in which an autonomous quadrotor must land on a moving platform. PPO is used to train two policies: RL and DR-RL. The former is used as a control policy, as well as in the outer-loop RL algorithm for MRAC-RL, while the latter is trained using a domain randomized training environment. We then compare MRAC-RL to these two standard RL techniques \cite{hodge2020deep}\cite{loquercio2020drone}.

\subsection{Quadrotor Dynamics}
We first describe the quadrotor dynamical model. The squared angular velocities of each propeller are used as input, that is: $u = [\omega_1^2, \omega_2^2, \omega_3^2, \omega_4^2]^T$. The thrust produced by each propeller is calculated by multiplying the squared angular speed by a propeller specific constant $\kappa$. Letting $K = diag(\kappa_1, \kappa_2, \kappa_3, \kappa_4)$, the vector of thrusts produced by each propeller is then given by $Ku$. Finally, denoting the (body-frame) vertical force, roll moment, pitch moment, and yaw moment by $f_z, \tau_\phi, \tau_\theta, \tau_\psi$ respectively, we have:

\begin{equation}\label{eq:qr_moment_dynamics}
\begin{aligned}
\begin{bmatrix}
f_z\\
\tau_\phi\\
\tau_\theta\\
\tau_\psi\\
\end{bmatrix} = 
\begin{bmatrix}
1 & 1 & 1 & 1\\
L & 0 & -L & 0\\
0 & L & 0 & -L\\
\mu & -\mu & \mu & -\mu\\
\end{bmatrix}Ku
\end{aligned}
\end{equation}
where $L$ is the distance from each propeller to the quadrotor center of mass, and $\mu$ is a rotational drag constant. Assuming low-speeds, we may then construct a simple rigid-body model for the quadrotor dynamics:
\begin{equation}\label{eq:nonlinear_dynamics}
\begin{split}
& \ddot{x} = (\cos \phi \cos \theta \cos \psi + \sin \phi \sin \psi )\frac{f_z}{m}\\
& \ddot{y} = (\cos \phi \sin \theta \sin \psi - \sin \phi \cos \psi)\frac{f_z}{m}\\
& \ddot{z} = \cos \phi \cos \theta \frac{f_z}{m} - g\\
& \ddot{\phi} = \dot{\theta}\dot{\psi}(\frac{I_y - I_z}{I_x}) + \frac{L}{I_x}\tau_\phi\\
& \ddot{\theta} = \dot{\phi}\dot{\psi}(\frac{I_z - I_x}{I_y}) + \frac{L}{I_y}\tau_\theta\\
& \ddot{\psi} = \dot{\phi}\dot{\theta}(\frac{I_x - I_y}{I_z}) + \frac{1}{I_z}\tau_\psi
\end{split}
\end{equation}

\noindent where $x,\; y,\; z$ represent the center of mass position in an inertial frame and $\phi, \theta, \psi$ are the roll, pitch, and yaw angles of the quadrotor body frame, respectively, in the inertial frame \cite{dydek2012adaptive}. $m$ is the mass of the quadrotor; $I_x, I_y, I_z$ are the moments of inertia. A linearized model of (\ref{eq:nonlinear_dynamics}) around the hover equilibrium point is given by:
\begin{equation}\label{eq:linearized_dynamics}
\begin{split}
& \ddot{x} = g\theta \quad \ddot{\theta} = \frac{L}{I_y}\tau_\theta \\
& \ddot{y} = -g\phi \quad \ddot{\phi} = \frac{L}{I_x}\tau_\phi\\
& \ddot{z} = \frac{\Delta f_z}{m} \quad \ddot{\psi} = \frac{1}{I_z}\tau{\psi}
\end{split}
\end{equation}
where $\Delta f_z = f_z - mg$. (\ref{eq:linearized_dynamics}) contains four distinct subsystems, which include the evolution of $[x, \dot{x}, \theta, \dot{\theta}]$, $[y, \dot{y}, \phi, \dot{\phi}]$, $[z, \dot{z}]$, and $[\psi, \dot{\psi}]$. We utilize (\ref{eq:linearized_dynamics}) as a design-model for the MRAC-RL controller, and (\ref{eq:nonlinear_dynamics}) as an evaluation and simulation model for the numerical experiment.

The model in (\ref{eq:nonlinear_dynamics}) is implemented using an RK4 integration scheme with a time step of $1$ millisecond. The following parameters are used to construct the reference environment: $I_x = I_y = .22 kg \cdot m^2, \; I_z = .44 kg \cdot m^2, \; m = 1.2kg, \; L = .30m$. $T_{max} = 20s$ in (\ref{eq:costfnc}). The control input is updated every $50$ milliseconds to emulate latencies due to measurement, communication, actuation and computation.

\subsection{Quadrotor Landing Task}
Due to the strength of RL algorithms in solving "unconstrained" control problems, we are free to define the landing task in an elegant and concise manner (instead of having to formulate a convex/simplified objective to enable tractable solutions via LQR or MPC). We assume that a platform of known inertial position is moving with known inertial velocity. The goal is to utilize full state feedback to determine a control policy that enables a quadrotor to landing on the moving platform from a wide array of initial conditions. Let $\Delta z$ and $\Delta xy$ be the inertial vertical distance and lateral distance, respectively, from the quadrotor to the platform. Furthermore, let $v_{xy} = \sqrt{\dot{x}^2 + \dot{y}^2}$ be the quadrotor's lateral velocity. We define the boolean variable $box$ to be True if ALL of the following simultaneously hold: $|\Delta z| \leq z_{max}$; $|\Delta xy| \leq l_{max}$; $|\phi| \leq \phi_{max}$; $|\theta| \leq \theta_{max}$; $|v_{xy}| \leq v_{l, max}$; $|v_z| \leq v_{z, max}$, where $z_{max}, l_{max}, \phi_{max}, \theta_{max}, v_{l, max}, v_{z, max}$ are user provided parameters that determine if the quadcopter has successfully landed.

A ternary cost function then quite naturally defines the control objective:
\begin{equation}\label{eq:costfnc}
\begin{aligned}
c(\vec{x}, t) = 
\begin{cases} 
      -1 & \textrm{IF } box \\
      1 & \textrm{ELSE IF } \Delta z \leq 0 \textrm{ OR } t \geq T_{max}\\
      0 & \textrm{ELSE} 
   \end{cases}
\end{aligned}
\end{equation}

\noindent where $\vec{x}$ captures the whole quadrotor state. The first case in (\ref{eq:costfnc}) defines success, the second case represents failure either due to a crash or a timeout, and the third is a neutral case and therefore set to zero. This cost function is a natural formulation for the problem at hand, as the goal is to have the quadrotor land as quickly and accurately as possible. Because the function is complex and non-quadratic, standard optimal control methods become inadequate.  RL is a good alternative as it allows the determination of a policy with such a cost function.

\subsection{Reinforcement Learning}
PPO is used to learn an appropriate feedback policy $\pi$ in order to solve the optimization problem in (\ref{eq:reward}). The actor and critic networks each have two hidden layers; each layer contains 64 neurons and uses tanh activation functions. A learning rate of $1$e$-4$, a discount factor of $.99$, a clipping range of $.2$, and a generalized advantage estimator discount of $.95$ are used as hyperparameters. Applying the PPO algorithm to the reference environment with the cost function in (\ref{eq:costfnc}) results in the successful training of a feedback policy $\pi$, which accomplishes the task when no parametric uncertainties or loss of propeller effectiveness are present.

\subsection{MRAC-RL}
We introduce two types of parametric uncertainties into the problem. In the first, the parameters $m$, $I_x$, $I_y$, $I_z$, and $L$ vary by $\pm25 \%$ of their reference values. In the second type, we test uncertainty in the form of a sudden and asymmetric loss of effectiveness (LOE) in one of the propellers. This is done by allowing $K$ in (\ref{eq:qr_moment_dynamics}) to take the form: $K = diag(\kappa_1, \kappa_2, \kappa_3, \beta \kappa_4)$, where $0 < \beta < 1$. This could simulate, for example, the sudden fracturing of a single propeller blade mid-flight, as was done in \cite{dydek2012adaptive}. This represents an extreme and sudden change to the quadrotor model that may occur in real-time. The goal is to determine an online control signal $u$ in (\ref{eq:qr_moment_dynamics})-(\ref{eq:nonlinear_dynamics}) so that the cost function in (\ref{eq:costfnc}) is minimized.

\subsection{DR-RL}\label{sec:drrl}
A standard RL method for accommodating model uncertainty is domain randomization (DR) \cite{muratore2018domain}\cite{loquercio2020drone}. In DR the agent is trained over a distribution of different environment models so that more robust and generalizable policies may be learned. Thus, we develop a domain-randomized RL (DR-RL) agent to compare against MRAC-RL. The DR-RL agent is trained using PPO in the same manner as the baseline RL agent, except that the environment parameters $m$, $I_x$, $I_y$, $I_z$, and $L$ are varied during training by sampling the parameter values from a uniform distribution over $\pm25 \%$ of the nominal values.

\subsection{Simulation Results}
When there is no parametric uncertainty or LOE, the aggregate success rate of the PPO-trained policies on the landing task was found to be $94\%$. This demonstrates that the PPO-trained policy (henceforth referred to as \textbf{\textit{RL}} below) is satisfactory when the target system and reference system are equivalent. We compare three control approaches:
\begin{itemize}
    \item \textbf{\textit{RL:}} The PPO-trained policy $\pi$ is used directly in the target system (as in Figure 1a). Every $50ms$ the quadrotor control is updated: $u = \pi(\vec{x})$, where $\vec{x}$ is the full state of the quadrotor.
    
    \item \textbf{\textit{MRAC-RL:}} The MRAC inner-loop converts $u_r$ to $u$ at each timestep (as in Figure 1b). Every $50ms$ the reference control is calculated: $u_r = \pi(\vec{x}_r)$. The quadrotor control $u$ is then calculated using (\ref{eq:xi}) - (\ref{eq:adaptive2}). 
    
    \item \textbf{\textit{DR-RL:}} 
    The DR-RL trained policy (described in section \ref{sec:drrl})) is used in the same manner as the \textbf{\textit{RL}} approach.
\end{itemize}

\begin{table}[h!]
  \begin{center}
    \caption{$\pm 25\%$ parametric uncertainty results}
    \begin{tabular}{c |c  c}
      \toprule % <-- Toprule here
      \multicolumn{1}{c|}{\textbf{Algorithm}} & \multicolumn{2}{c|}{\textbf{Results}}\\
      \toprule
      \textbf{} &  \textbf{Success Rate} & 
      \textbf{Avg. Success Time}\\

      \midrule % <-- Midrule here
      \textit{RL} &            $48\%$ &         $7.5s$\\
      \textit{MRAC-RL} & $        \mathbf{82}\%$ & $\mathbf{3.5s}$\\
      \textit{DR-RL} &            $74\%$ &         $8.9s$\\

      \bottomrule % <-- Bottomrule here
    \end{tabular}
      \label{tab:table1}
   
  \end{center}

\end{table}
As shown in Table 1, the inclusion of an MRAC inner-loop increases the task success rate when compared to both standard RL and DR-RL. The latter improvement is noteworthy, as DR-RL is explicitly used to accomodate variations in model parameters. 

We next compare the PPO-trained policy to MRAC-RL in the case of a sudden asymmetric loss of effectiveness (LOE) - that is, when the ability of a single propeller to produce thrust is severely compromised. 

\begin{table}[h!]
  \begin{center}
  
     \caption{Asymmetric LOE. The LOE column is the degree of propeller thrust lost (with $0\%$ being no loss).}
    \begin{tabular}{c| c |c}
      \toprule % <-- Toprule here
      \textbf{PPO} & \textbf{MRAC-RL} & \textbf{LOE}\\
        \textbf{Success Rate} & \textbf{Success Rate} & \\
      \midrule %
      
      $94\%$ & $--$ & $0\%$\\
      $71\%$ & $95\%$ & $10\%$\\
      $28\%$ & $81\%$ & $25\%$\\
      $4\%$ & $47\%$ & $50\%$\\
      $0\%$ & $11\%$ & $75\%$\\

    \end{tabular}
    
    \label{tab:results2}
\end{center}
\end{table}
\begin{wrapfigure}{L}{0.24\textwidth}
\centering
\includegraphics[width=0.24\textwidth]{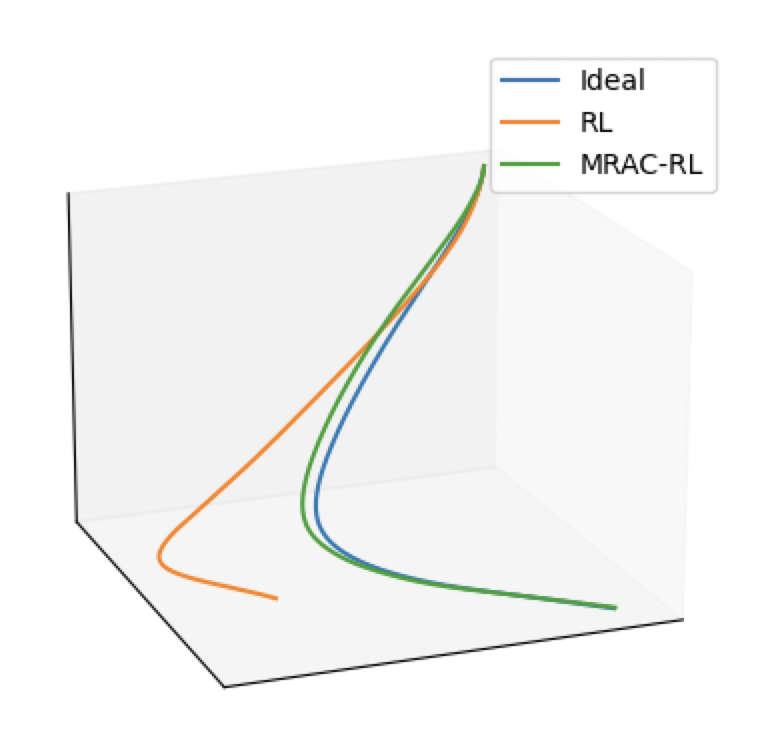}
\caption{\label{fig:plot}MRAC-RL (green) and pure RL (orange) trajectory rollouts when a single propeller loses effectiveness. The baseline "ideal" trajectory (blue) is shown, when there is no loss of effectiveness. Note that the MRAC-RL trajectory more closely tracks this baseline than the RL rollout.}
\end{wrapfigure}
As in the case of simple static model perturbations, the inclusion of MRAC at the inner-loop greatly improves the task performance. In the LOE experiment, we determine the state trajectories produced by RL and MRAC-RL algorithms, and compare both trajectories with the ideal baseline trajectory when there is no LOE. We report that, for a LOE of $10\%$, the standard RL approach exhibits an average point-wise trajectory divergence of $.17m$, while the MRAC-RL approach leads to a divergence of $.01m$. For a LOE of $50\%$ the RL divergence is $.47m$ and the MRAC-RL divergence is $.04m$.

\section{SUMMARY \& CONCLUSIONS}\label{sec: 5}
In this paper we have proposed a control architecture, MRAC-RL, to solve optimal control problems in the form of (\ref{eq:PCC})-(\ref{eq:optimal_control}) when parametric modeling uncertainty is present. MRAC-RL consists of an MRAC-based inner-loop and an RL-based outer-loop. We provide stability guarantees for the resulting closed-loop system and prove that the adaptive tracking objective is achieved. An extensive numerical investigation of a quadrotor landing task is carried out to validate the proposed MRAC-RL. Both parametric uncertainties and an asymmetric loss of actuator effectiveness are emulated in this numerical study. In addition to validating the theoretical result of satisfactory tracking in the presence of modeling errors, we also show that the performance of the controller can be improved by the inclusion of MRAC. This improvement is quantified through a comparison with other approaches that are based only on RL. While this paper compares MRAC to standard RL and domain-randomization techniques, the modular nature of the proposed architecture allows integration with several other robust RL methods. For example, a robust algorithm such as PILCO \cite{Deisenroth2011} could be used to train the outer-loop.

In this paper we have paid special attention to the minimization and convergence of \textit{tracking} error, but did not address the convergence of \textit{parameter} error, as our focus is on control performance with imperfect learning. Incorporation of persistent excitation conditions to investigate improved learning, as was done in in \cite{dean2019safely} and \cite{narendra1989}, is the subject of future research. Proof of control optimality under general conditions of finite time-horizons and arbitrary costs, and experimental demonstrations of MRAC-RL are all topics for future research as well. 

%%%%%%%%%%%%%%%%%%%%%%%%%%%%%%%%%%%%%%%%%%%%%%%%%%%%%%%%%%%%%%%%%%%%%%%%%%%%%%%%

\bibliographystyle{IEEEtran}
\bibliography{references}

\begin{thebibliography}{10}
\providecommand{\url}[1]{#1}
\csname url@rmstyle\endcsname
\providecommand{\newblock}{\relax}
\providecommand{\bibinfo}[2]{#2}
\providecommand\BIBentrySTDinterwordspacing{\spaceskip=0pt\relax}
\providecommand\BIBentryALTinterwordstretchfactor{4}
\providecommand\BIBentryALTinterwordspacing{\spaceskip=\fontdimen2\font plus
\BIBentryALTinterwordstretchfactor\fontdimen3\font minus
  \fontdimen4\font\relax}
\providecommand\BIBforeignlanguage[2]{{%
\expandafter\ifx\csname l@#1\endcsname\relax
\typeout{** WARNING: IEEEtran.bst: No hyphenation pattern has been}%
\typeout{** loaded for the language `#1'. Using the pattern for}%
\typeout{** the default language instead.}%
\else
\language=\csname l@#1\endcsname
\fi
#2}}

\bibitem{mnih2015}
V.~Mnih, K.~Kavukcuoglu, D.~Silver, A.~A. Rusu, J.~Veness, M.~G. Bellemare,
  A.~Graves, M.~Riedmiller, A.~K. Fidjeland, G.~Ostrovski, \emph{et~al.},
  ``Human-level control through deep reinforcement learning,'' \emph{nature},
  vol. 518, no. 7540, pp. 529--533, 2015.

\bibitem{Ng2006}
A.~Y. Ng, A.~Coates, M.~Diel, V.~Ganapathi, J.~Schulte, B.~Tse, E.~Berger, and
  E.~Liang, ``Autonomous inverted helicopter flight via reinforcement
  learning,'' in \emph{Experimental robotics IX}.\hskip 1em plus 0.5em minus
  0.4em\relax Springer, 2006, pp. 363--372.

\bibitem{lillicrap2019continuous}
T.~P. Lillicrap, J.~J. Hunt, A.~Pritzel, N.~Heess, T.~Erez, Y.~Tassa,
  D.~Silver, and D.~Wierstra, ``Continuous control with deep reinforcement
  learning,'' \emph{arXiv preprint arXiv:1509.02971}, 2015.

\bibitem{kober2013}
J.~Kober, J.~A. Bagnell, and J.~Peters, ``Reinforcement learning in robotics: A
  survey,'' \emph{The International Journal of Robotics Research}, vol.~32,
  no.~11, pp. 1238--1274, 2013.

\bibitem{recht2018tour}
B.~Recht, ``A tour of reinforcement learning: The view from continuous
  control,'' \emph{Annual Review of Control, Robotics, and Autonomous Systems},
  vol.~2, pp. 253--279, 2019.

\bibitem{schulman2018highdimensional}
J.~Schulman, P.~Moritz, S.~Levine, M.~Jordan, and P.~Abbeel, ``High-dimensional
  continuous control using generalized advantage estimation,'' \emph{arXiv
  preprint arXiv:1506.02438}, 2015.

\bibitem{Koos2010}
S.~Koos, J.-B. Mouret, and S.~Doncieux, ``Crossing the reality gap in
  evolutionary robotics by promoting transferable controllers,'' in
  \emph{Proceedings of the 12th annual conference on Genetic and evolutionary
  computation}, 2010, pp. 119--126.

\bibitem{tan2018simtoreal}
J.~Tan, T.~Zhang, E.~Coumans, A.~Iscen, Y.~Bai, D.~Hafner, S.~Bohez, and
  V.~Vanhoucke, ``Sim-to-real: Learning agile locomotion for quadruped
  robots,'' \emph{arXiv preprint arXiv:1804.10332}, 2018.

\bibitem{fulton2018safe}
N.~Fulton and A.~Platzer, ``Safe reinforcement learning via formal methods,''
  in \emph{AAAI Conference on Artificial Intelligence}, 2018.

\bibitem{AurkoMismatch}
A.~Roy, H.~Xu, and S.~Pokutta, ``Reinforcement learning under model mismatch,''
  in \emph{Advances in neural information processing systems}, 2017, pp.
  3043--3052.

\bibitem{rajeswaran2018generalization}
A.~Rajeswaran, K.~Lowrey, E.~V. Todorov, and S.~M. Kakade, ``Towards
  generalization and simplicity in continuous control,'' in \emph{Advances in
  Neural Information Processing Systems}, 2017, pp. 6550--6561.

\bibitem{packer2019assessing}
C.~Packer, K.~Gao, J.~Kos, P.~Kr{\"a}henb{\"u}hl, V.~Koltun, and D.~Song,
  ``Assessing generalization in deep reinforcement learning,'' \emph{arXiv
  preprint arXiv:1810.12282}, 2018.

\bibitem{schulman2017proximal}
J.~Schulman, F.~Wolski, P.~Dhariwal, A.~Radford, and O.~Klimov, ``Proximal
  policy optimization algorithms,'' \emph{arXiv preprint arXiv:1707.06347},
  2017.

\bibitem{wang2019benchmarking}
E.~Langlois, S.~Zhang, G.~Zhang, P.~Abbeel, and J.~Ba, ``Benchmarking
  model-based reinforcement learning,'' \emph{arXiv preprint arXiv:1907.02057},
  2019.

\bibitem{duan2016benchmarking}
Y.~Duan, X.~Chen, R.~Houthooft, J.~Schulman, and P.~Abbeel, ``Benchmarking deep
  reinforcement learning for continuous control,'' in \emph{International
  Conference on Machine Learning}, 2016, pp. 1329--1338.

\bibitem{henderson2019deep}
P.~Henderson, R.~Islam, P.~Bachman, J.~Pineau, D.~Precup, and D.~Meger, ``Deep
  reinforcement learning that matters,'' \emph{arXiv preprint
  arXiv:1709.06560}, 2017.

\bibitem{higgins2018darla}
I.~Higgins, A.~Pal, A.~A. Rusu, L.~Matthey, C.~P. Burgess, A.~Pritzel,
  M.~Botvinick, C.~Blundell, and A.~Lerchner, ``{DARLA}: Improving zero-shot
  transfer in reinforcement learning,'' \emph{arXiv preprint arXiv:1707.08475},
  2017.

\bibitem{nagabandi2019learning}
A.~Nagabandi, I.~Clavera, S.~Liu, R.~S. Fearing, P.~Abbeel, S.~Levine, and
  C.~Finn, ``Learning to adapt in dynamic, real-world environments through
  meta-reinforcement learning,'' \emph{arXiv preprint arXiv:1803.11347}, 2018.

\bibitem{lake2016building}
B.~M. Lake, T.~D. Ullman, J.~B. Tenenbaum, and S.~J. Gershman, ``Building
  machines that learn and think like people,'' \emph{Behavioral and brain
  sciences}, vol.~40, 2017.

\bibitem{Deisenroth2011}
M.~Deisenroth and C.~E. Rasmussen, ``{PILCO}: A model-based and data-efficient
  approach to policy search,'' in \emph{Proceedings of the 28th International
  Conference on machine learning (ICML-11)}, 2011, pp. 465--472.

\bibitem{rajeswaran2017epopt}
A.~Rajeswaran, S.~Ghotra, B.~Ravindran, and S.~Levine, ``Epopt: Learning robust
  neural network policies using model ensembles,'' \emph{arXiv preprint
  arXiv:1610.01283}, 2016.

\bibitem{loquercio2020drone}
A.~Loquercio, E.~Kaufmann, R.~Ranftl, A.~Dosovitskiy, V.~Koltun, and
  D.~Scaramuzza, ``Deep drone racing: From simulation to reality with domain
  randomization,'' \emph{IEEE Transactions on Robotics}, vol.~36, no.~1, pp.
  1--14, 2019.

\bibitem{pinto2017robust}
L.~Pinto, J.~Davidson, R.~Sukthankar, and A.~Gupta, ``Robust adversarial
  reinforcement learning,'' \emph{arXiv preprint arXiv:1703.02702}, 2017.

\bibitem{berkenkamp2017safe}
F.~Berkenkamp, M.~Turchetta, A.~Schoellig, and A.~Krause, ``Safe model-based
  reinforcement learning with stability guarantees,'' in \emph{Advances in
  neural information processing systems}, 2017, pp. 908--918.

\bibitem{L1AC}
T.~Leman, E.~Xargay, G.~Dullerud, N.~Hovakimyan, and T.~Wendel, ``L1 adaptive
  control augmentation system for the x-48b aircraft,'' in \emph{AIAA guidance,
  navigation, and control conference}, 2009, p. 5619.

\bibitem{dydek2013}
Z.~T. Dydek, A.~M. Annaswamy, and E.~Lavretsky, ``Adaptive control of quadrotor
  uavs: A design trade study with flight evaluations,'' \emph{IEEE Transactions
  on control systems technology}, vol.~21, no.~4, pp. 1400--1406, 2012.

\bibitem{WISE2011}
K.~A. Wise and E.~Lavretsky, ``Robust and adaptive control of x-45a j-ucas: a
  design trade study,'' \emph{IFAC Proceedings Volumes}, vol.~44, no.~1, pp.
  6555--6560, 2011.

\bibitem{michini2009}
B.~Michini and J.~How, ``L1 adaptive control for indoor autonomous vehicles:
  Design process and flight testing,'' in \emph{AIAA Guidance, Navigation, and
  Control Conference}, 2009, p. 5754.

\bibitem{Wiese2013}
D.~P. Wiese, A.~M. Annaswamy, J.~A. Muse, and M.~A. Bolender, ``Adaptive
  control of a generic hypersonic vehicle,'' in \emph{AIAA Guidance,
  Navigation, and Control (GNC) Conference}, 2013, p. 4514.

\bibitem{narendra1989}
K.~S. {Narendra} and A.~M. {Annaswamy}, \emph{Stable Adaptive Systems}.\hskip
  1em plus 0.5em minus 0.4em\relax Reprinted 2004, Dover Publications, 1989.

\bibitem{aastrom2013adaptive}
K.~J. {\AA}str{\"o}m and B.~Wittenmark, \emph{Adaptive control}.\hskip 1em plus
  0.5em minus 0.4em\relax Courier Corporation, 2013.

\bibitem{ioannou2012robust}
P.~A. Ioannou and J.~Sun, \emph{Robust adaptive control}.\hskip 1em plus 0.5em
  minus 0.4em\relax Courier Corporation, 2012.

\bibitem{Karason1994}
S.~P. {K\'arason} and A.~M. {Annaswamy}, ``Adaptive control in the presence of
  input constraints,'' \emph{IEEE Transactions on Automatic Control}, vol.~39,
  no.~11, pp. 2325--2330, 1994.

\bibitem{Lavretsky2004}
E.~Lavretsky and N.~Hovakimyan, ``Positive/spl mu/-modification for stable
  adaptation in the presence of input constraints,'' in \emph{Proceedings of
  the 2004 American Control Conference}, vol.~3.\hskip 1em plus 0.5em minus
  0.4em\relax IEEE, 2004, pp. 2545--2550.

\bibitem{gaudio2019adaptive}
J.~E. Gaudio, A.~M. Annaswamy, M.~A. Bolender, and E.~Lavretsky, ``Adaptive
  flight control in the presence of limits on magnitude and rate,'' \emph{arXiv
  preprint arXiv:1907.11913}, 2019.

\bibitem{sutton}
R.~S. Sutton, A.~G. Barto, and R.~J. Williams, ``Reinforcement learning is
  direct adaptive optimal control,'' \emph{IEEE Control Systems Magazine},
  vol.~12, no.~2, pp. 19--22, 1992.

\bibitem{kaelbling1996reinforcement}
L.~P. Kaelbling, M.~L. Littman, and A.~W. Moore, ``Reinforcement learning: A
  survey,'' \emph{Journal of artificial intelligence research}, vol.~4, pp.
  237--285, 1996.

\bibitem{schulman2015high}
J.~Schulman, P.~Moritz, S.~Levine, M.~Jordan, and P.~Abbeel, ``High-dimensional
  continuous control using generalized advantage estimation,'' \emph{arXiv
  preprint arXiv:1506.02438}, 2015.

\bibitem{vidyasagar2020recent}
M.~Vidyasagar, ``Recent advances in reinforcement learning,'' in \emph{2020
  American Control Conference (ACC)}.\hskip 1em plus 0.5em minus 0.4em\relax
  IEEE, 2020, pp. 4751--4756.

\bibitem{watkins1992q}
C.~J. Watkins and P.~Dayan, ``Q-learning,'' \emph{Machine learning}, vol.~8,
  no. 3-4, pp. 279--292, 1992.

\bibitem{devraj2017zap}
A.~M. Devraj and S.~P. Meyn, ``Zap q-learning,'' in \emph{Proceedings of the
  31st International Conference on Neural Information Processing Systems},
  2017, pp. 2232--2241.

\bibitem{hodge2020deep}
V.~J. Hodge, R.~Hawkins, and R.~Alexander, ``Deep reinforcement learning for
  drone navigation using sensor data,'' \emph{Neural Computing and
  Applications}, pp. 1--19, 2020.

\bibitem{dydek2012adaptive}
Z.~T. Dydek, A.~M. Annaswamy, and E.~Lavretsky, ``Adaptive control of quadrotor
  uavs: A design trade study with flight evaluations,'' \emph{IEEE Transactions
  on control systems technology}, vol.~21, no.~4, pp. 1400--1406, 2012.

\bibitem{muratore2018domain}
F.~Muratore, F.~Treede, M.~Gienger, and J.~Peters, ``Domain randomization for
  simulation-based policy optimization with transferability assessment,'' in
  \emph{Conference on Robot Learning}.\hskip 1em plus 0.5em minus 0.4em\relax
  PMLR, 2018, pp. 700--713.

\bibitem{dean2019safely}
S.~Dean, S.~Tu, N.~Matni, and B.~Recht, ``Safely learning to control the
  constrained linear quadratic regulator,'' 2019.

\end{thebibliography}

\end{document}